\documentclass[aps,prl,twocolumn,showpacs,superscriptaddress,floatfix]{revtex4-1}
\usepackage{dcolumn}
\usepackage{amsmath}
\usepackage{graphicx,epsfig,psfrag}
\usepackage{amssymb}
\usepackage{natbib}
\usepackage{url}
\usepackage[breaklinks=true]{hyperref}
\usepackage{mathtools}
\usepackage{subfigure}
\hypersetup{
        colorlinks=true,
}
\usepackage{bookmark}
\usepackage{epic}

\usepackage{txfonts}
\usepackage{color}

\begin{document}
\title{Magnetic field dependence of the thermopower of Kondo-correlated quantum dots}
\author{T. A. Costi}
\affiliation 
{Peter Gr\"{u}nberg Institut and Institute for Advanced Simulation, 
Research Centre J\"ulich, 52425 J\"ulich, Germany}
\begin{abstract}
Recent experiments have measured the signatures of the Kondo effect in the zero-field thermopower of strongly
correlated quantum dots [Svilans {\em et al.,} Phys. Rev. Lett. {\bf 121}, 206801 (2018); Dutta {\em et al.,} Nano Lett. {\bf 19}, 506 (2019)]. They confirm the predicted Kondo-induced sign change in the thermopower, upon increasing the temperature through a gate-voltage dependent value $T_{1}\gtrsim T_{\rm K}$, where $T_{\rm K}$ is the Kondo temperature. Here, we use the numerical renormalization group (NRG) method to investigate the effect of a finite magnetic field $B$ on the thermopower of such quantum dots. We show that, for fields $B$ exceeding a gate-voltage dependent value $B_{0}$,
an additional sign change takes place in the Kondo regime at a temperature $T_{0}(B\geq B_{0})>0$ with $T_0<T_1$. 
The field $B_{0}$ is comparable to, but larger than, the field $B_{c}$ at which the zero-temperature spectral function splits in a magnetic field. The validity of the NRG results for $B_{0}$ are checked by comparison with asymptotically exact higher-order Fermi-liquid calculations [Oguri {\em et al.,} Phys. Rev. B {\bf 97}, 035435 (2018)].
Our calculations clarify the field-dependent signatures of the Kondo effect in the thermopower of Kondo-correlated quantum dots and explain the recently measured trends in the $B$-field dependence of the thermoelectric response of such systems
[Svilans {\em et al.,} Phys. Rev. Lett. {\bf 121}, 206801 (2018)].
\end{abstract}
\maketitle
{\em Introduction.}\,Understanding the thermoelectric transport through gate-tunable Kondo-correlated molecules \cite{Park2002,Yu2005,Scott2010}, adatoms \cite{Li1998,Madhavan1998,Manoharan2000,Nagaoka2002,Wahl2004,Ternes2017} , and semiconductor quantum dots  \cite{Goldhaber1998b,Cronenwett1998,Schmid1998,vanderWiel2000,Kretinin2011} poses both experimental and theoretical challenges \cite{Sothmann2014,Zimbovskaya2016,Thoss2018}. While electrical conductance measurements on nanosystems are standard, applying a quantifiable temperature gradient, and measuring the resulting thermovoltage, across such small systems is experimentally challenging \cite{Scheibner2005,Reddy2007,Cui2017,Prete2019}. On the theoretical side a description of the Kondo-induced transport at low temperatures \cite{Ng1988,Glazman1988,Pustilnik2004}, requires, in general, the use of non-perturbative methods \cite{Wilson1975,KWW1980a,Gonzalez-Buxton1998,Bulla2008,Tsvelick1983b,Andrei2013,Gull2011b,White1992}, and, moreover, the above systems can be routinely driven out of equilibrium \cite{DeFranceschi2002, Leturcq2005,Josefsson2018}, thus posing additional challenges, such as the description of dissipative processes in the nonequilibrium Kondo effect \cite{Hershfield1991,Hershfield1993,Meir1993,Koenig1996,Rosch2003a,Anders2008a,Mehta2008,Moca2011,Pletyukhov2012,Munoz2013,Fugger2018,Schwarz2018,Oguri2018a}. In this Rapid Communication, motivated by a recent experiment \cite{Svilans2018}, we focus on the magnetic field dependence of the thermopower of Kondo-correlated quantum dots, which, unlike the magnetoconductance \cite{Costi2001,Hofstetter2001,Karrasch2006}, has received almost no theoretical attention \cite{Sakano2007,Weymann2013,Andergassen2011a,Rejec2012}, although it also exhibits, as we shall show,
marked signatures of the Kondo effect.  

Specifically, we focus on a gate-tunable quantum dot  described by an Anderson impurity model. In  zero magnetic field, both its thermopower, $S(T)$,  and  electrical conductance, $G(T)$, have been thoroughly investigated as a function of temperature and gate voltage, and, characteristic signatures of the Kondo effect have been identified in both  $G(T)$ \cite{Glazman1988,Ng1988,Costi1994,Pustilnik2004} and $S(T)$ \cite{Costi2010}. 
A hallmark of the Kondo effect in $G(T)$ is the lifting of the Coulomb blockade at low temperatures $T$ below the Kondo scale $T_{\rm K}$ \cite{Glazman1988,Ng1988,Costi1994,Pustilnik2004}, which has been verified in many experiments \cite{Goldhaber1998b,Cronenwett1998, Schmid1998,vanderWiel2000,Kretinin2011}, and reflects the development of the Kondo resonance at the Fermi level upon decreasing temperature. Signatures of the Kondo effect in the zero-field thermopower are more subtle, since the thermopower probes the
asymmetry of the Kondo resonance about the Fermi level, and thus, its sign reflects the relative importance, at any given temperature $T$, of electron- or holelike contributions to the transport integrals in the definition of $S(T)$.
It has been found theoretically \cite{Costi2010}, and verified in recent experiments \cite{Svilans2018,Dutta2019}, that a hallmark of the Kondo effect in the zero-field thermopower, is a sign change at a characteristic temperature $T_1\gtrsim T_{\rm K}$, which is absent in the other regimes \cite{Costi2010,Dutta2019}. Given the above, and since it is well known that a magnetic field has a large effect on the Kondo resonance \cite{Costi2000}, the question we ask in this Rapid Communication is whether a magnetic field gives rise to additional characteristic signatures of the Kondo effect in the thermopower of quantum dots ? We show that this is the case and provide an interpretation of the recent experiment of  Svilans {\em et al.}~\onlinecite{Svilans2018}.

{\em Model and transport calculations.}\,We describe the thermoelectric transport through a strongly correlated quantum dot within a single level Anderson impurity model, $H=H_{\rm dot} + H_{\rm leads} + H_{\rm tunneling}$. Here, $H_{\rm dot}=\sum_{\sigma}\varepsilon_{0}n_{0\sigma}-g\mu_{\rm B}BS_{z}+Un_{0\uparrow}n_{0\downarrow}$, describes the quantum dot with energy level $\varepsilon_{0}$ and
local Coulomb repulsion $U$ in a magnetic field $B$, with $S_{z}=\frac{1}{2}(n_{0\uparrow}-n_{0\downarrow})$.  $H_{\rm leads}=\sum_{k\alpha=L,R\sigma}\epsilon_{k\alpha}c^{\dagger}_{k\alpha\sigma}c_{k\alpha\sigma}$
describes conduction electron leads ($\alpha=L,R$), with kinetic energies $\epsilon_{k\alpha}$, and $H_{\rm tunneling}=\sum_{k\alpha\sigma} t_{\alpha}(c^{\dagger}_{k\alpha\sigma}d_{\sigma}+d^{\dagger}_{\sigma}c_{k\alpha\sigma})$ describes the tunneling
of electrons from the leads to the dot with amplitudes $t_{\alpha=L,R}$. In the above, $n_{0\sigma}=d_{\sigma}^{\dagger}d_{\sigma}$ is the number operator for electrons on the dot, $d_{\sigma}^{\dagger}$ ($d_{\sigma}$) and
$c_{k\alpha\sigma}^{\dagger}$ ($c_{k\alpha\sigma}$ ) are electron creation (annihilation) operators, and we assume 
a constant density of states, $\rho_{\alpha}(\omega)=\sum_{k}\delta(\omega-\varepsilon_{k\alpha})=1/(2D) \equiv N_{\rm F}$ for both leads, with $D=1$ the half-bandwidth. The strength of correlations is characterized by $U/\Gamma$, where $\Gamma=2\pi N_{\rm F}(t_L^2 +t_R^2)$ is the tunneling rate, taken throughout as $\Gamma=0.002D$.
We solve $H$ using the numerical renormalization group (NRG) technique \cite{Wilson1975,KWW1980a,Gonzalez-Buxton1998,Bulla2008,Zitko2009b},
and exemplify results primarily for  $U/\Gamma=8$, or for $U/\Gamma=3.2$, relevant to the experiment \cite{Svilans2018}. With the dimensionless gate voltage ${\rm v}_g\equiv (\varepsilon_{0}+U/2)/\Gamma$, the particle-hole symmetric (or midvalley) point $\varepsilon_{0}=-U/2$, where $n_{0}=\sum_{\sigma}n_{0\sigma}=1$, occurs at ${\rm v}_g=0$. The Kondo scale, $T_{\rm K}$, is obtained from the $T=B=0$ spin susceptibility $\chi_0$ via $\chi_0=g\mu_{\rm B}/4k_{\rm B} T_{\rm K}$, and is comparable to $k_{\rm B}T_{\rm K1}/\Gamma=\sqrt{U/4\Gamma}\exp(-\pi|\varepsilon_0||\varepsilon_0+U|/\Gamma U)$ from perturbative scaling \cite{Haldane1978,Hewson1993}. The thermopower $S(T)=-I_{1}/|e|TI_{0}$ \cite{Kim2002,Dong2002,Costi2010} is calculated by evaluating the transport integrals $I_{m=0,1}=\gamma\int_{-\infty}^{+\infty}d\omega (-\partial f/\partial \omega) \omega^m A(\omega,T)$, with $\gamma=\pi\Gamma/2h$, directly from the discrete Lehmann representation of  $A(\omega,T)=\sum_{\sigma}A_{\sigma}(\omega,T)$ \cite{Yoshida2009}, where $A_{\sigma}(\omega,T)$ is the spin-resolved local level spectral function. In the following, we focus on ${\rm v}_g>0$. Results for ${\rm v}_g<0$ follow by particle-hole symmetry: $S_{-{\rm v}_g}(T)=-S_{+{\rm v}_g}(T)$.

\begin{figure}[t]
\centering 
\includegraphics[width=0.95\columnwidth]{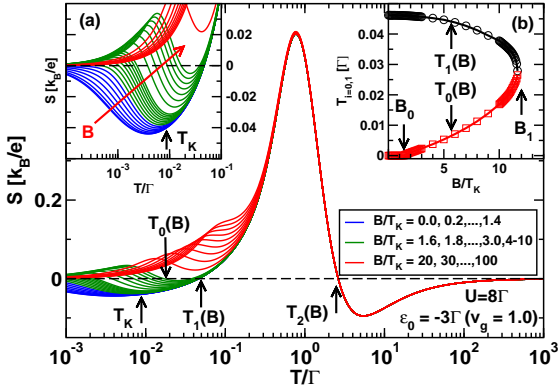}
\caption 
{Thermopower $S$ (in units of $k_{\rm B}/e=86.17{\,\rm \mu V/K}$) vs temperature $T/\Gamma$ of a Kondo-correlated quantum dot for increasing values of the magnetic field $B/T_{\rm K}$ [$U=8\Gamma$, $\varepsilon_{0}=-3\Gamma$ (${\rm v}_{g}=1.0$), $T_{\rm K}/\Gamma = 4.32\times 10^{-3}$]. 
 For $B<B_{0}\approx 1.45T_{\rm K}$ (blue solid lines), two sign changes are found at $T_{1}(B)$ and $T_{2}(B)$ , whereas for $B_{0}<B<B_{1}$ (green solid lines) an additional sign change occurs at a temperature $T_{0}(B)$, and, for $B>B_{1}$ (red solid lines), only the sign change at $T_{2}(B)$ is present. 
Inset (a): Evolution of the Kondo-induced thermopower peak with increasing $B$ (red arrow). 
Inset (b): $T_{0}(B)$ and $T_{1}(B)$ vs $B/T_{\rm K}$. $B_0$ and  $B_1$ are also indicated. 
 NRG parameters: discretization parameter $\Lambda=4$, $z$ averaging \cite{Oliveira1994,Campo2005} with $N_z=4$, retaining $N_{\rm states}=900$ states. $v_g$
}
\label{fig:fig1}
\end{figure}
{\em Thermopower in a magnetic field.}\,Figure~\ref{fig:fig1} shows the effect of a magnetic field on the temperature dependence of the thermopower for typical parameters in the Kondo regime
\footnote{See Ref.~\onlinecite{Costi2019b} for results outside the Kondo regime}.
It is useful to first briefly recapitulate the behavior of $S(T)$ at $B=0$ \cite{Costi2010} (also shown in Fig.~\ref{fig:fig1}):
for ${\rm v}_g>0$ the thermopower, $S(T)$, exhibits a (negative) Kondo-induced thermopower peak at $T\approx T_{\rm K}$ and
two sign changes at the gate-voltage dependent temperatures $T_{1}\gtrsim T_{\rm K}$ and $T_{2}\gtrsim \Gamma$, which
are characteristic of the Kondo regime, and, are absent in the other regimes, where $S(T)$ is of one sign
\cite{Costi2010}.  Unlike $T_{\rm K}$, neither $T_{1}$ nor $T_2$ are low-energy scales, since they are not exponentially small in $U/\Gamma$ \cite{Costi2010,Dutta2019}.
They are nevertheless closely connected to Kondo physics \cite{Costi2010,Dutta2019}. For example, the sign change at $T_1$ results from a rearrangement of spectral weight in the asymmetrically located Kondo resonance with increasing temperature \cite{Dutta2019}.
\begin{figure}[t]
\centering 
  \includegraphics[width=0.95\columnwidth]{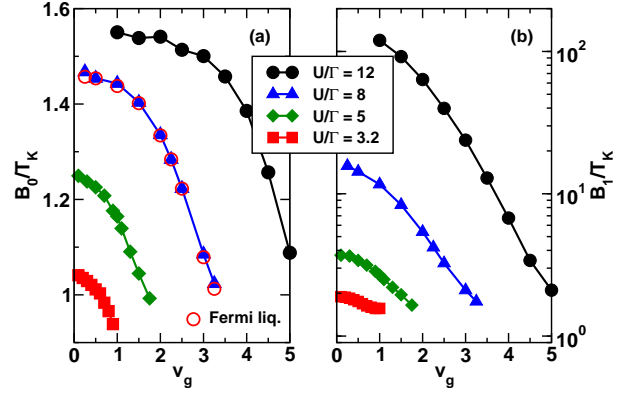}
\caption 
{NRG results (filled symbols) for, (a), $B_{0}/T_{\rm K}$ vs ${\rm v}_{g}$, and, (b), $B_{1}/T_{\rm K}$ vs ${\rm v}_{g}$ (in the Kondo regime),  for a range of $U/\Gamma$. 
Open circles in (a): Fermi-liquid result for $B_{0}/T_{\rm K}$ at $U/\Gamma=8$.  NRG parameters as in Fig.~\ref{fig:fig1}
}
\label{fig:fig2}
\end{figure}

For $B>0$, the above picture is modified as follows: initially, for low fields $B\lesssim T_{\rm K}$, the
thermopower $S(T)$ has a similar temperature dependence as for $B=0$, with two sign changes at $T_1(B)$ and $T_2(B)$,
where $T_{1}(B)$ and $T_{2}(B)$ are the finite-$B$ analogs of the two temperatures $T_1$ and $T_2$ where $S(T)$ changes sign at $B=0$.
The main effect of $B$ on $S(T)$ in this low-field limit is to shift the  Kondo-induced peak in $S(T)$ at $T\approx T_{\rm K}$ to
higher temperatures and to reduce it in amplitude with increasing $B$, while leaving its sign unchanged [see Fig.~\ref{fig:fig1}(a)],
a trend also seen experimentally in heavy fermion systems \footnote{H. von L{\"o}hneysen, private communication and \protect{\cite{Merker2016}}}.
Once $B$ exceeds a gate-voltage dependent value, $B_{0}$, the thermopower exhibits an additional sign change at a temperature
$T_{0}(B)<T_{1}(B)<T_{2}(B)$. While $T_{0}(B)$ and $T_{1}(B)$ have a significant $B$ dependence
[Fig.~\ref{fig:fig1}(b)], $T_{2}(B)$ (for the present parameters) is essentially $B$ independent (Fig.~\ref{fig:fig1}). Further increasing $B$
towards the gate-voltage dependent value $B_1$ results in a merging of $T_{0}(B)$ and $T_{1}(B)$ to a common value at $B=B_1$ [Fig.~\ref{fig:fig1}(b)]. For the parameters in Fig.~\ref{fig:fig2}, for example, we have
$T_{0}(B_1)=T_{1}(B_1)\approx 0.027\Gamma$ and $B_{1}\approx 12 T_{\rm K}\gg B_{0}\approx 1.45T_{\rm K}$.
For still larger $B$-values $B>B_1$ (and for ${\rm v}_g$ still in the Kondo regime), 
only the sign change at $T_2$ remains. Thus, in the Kondo regime, a sign change in $S(T)$ at  $T=T_{0}(B)$
for $B_0<B<B_1$ is an additional characteristic feature of the Kondo effect in $S(T)$.

Of particular interest for experiments are the magnitudes of  $B_{0}$ and $B_{1}$ for
quantum dots with different values of $U/\Gamma$. These are shown in 
Figs.~\ref{fig:fig2}(a) and \ref{fig:fig2}(b), respectively, for gate voltages in the Kondo regime. While $B_{0}$ is of order $T_{\rm K}$ for all values of $U/\Gamma$, $B_1$ is typically much larger, being of order $T_1$ for $U/\Gamma\gg 1$, and only approaching values of order $T_{\rm K}$ for smaller values of $U/\Gamma$ \cite{Costi2019b}: e.g., for the moderately correlated quantum dot of the experiment \cite{Svilans2018}, with $U/\Gamma=3.2$, we have that $1.5T_{\rm K}\lesssim B_{1}\lesssim 2.0 T_{\rm K}$ and $ 0.94T_{\rm K}\lesssim B_{0}\lesssim 1.04 T_{\rm K}$ for gate voltages in the Kondo regime. The value of $B_{0}$ can also be extracted from Fermi-liquid theory, since the opposite signs of the thermopower for $B<B_{0}$ and $B>B_{0}$ persist to asymptotically low temperatures $T\ll T_{\rm K}$ [Fig.~\ref{fig:fig1}(a)]. At such low temperatures, a Sommerfeld expansion for the thermopower gives
\begin{align}
  S(T) &\approx -\frac{k_{\rm B}}{|e|}\frac{\pi^2}{3}k_{\rm B}T \frac{1}{A(0,0)}\frac{\partial}{\partial\omega} A(\omega,T=0)|_{\omega=0},\label{eq:sommerfeld}
\end{align}
which also makes clear the physical significance of $B_0$ as the field where the slope of the spectral
  function at the Fermi level changes sign.
Making use of the Fermi-liquid expressions for $A(\omega,T)$ to leading order in $\omega$ and $T$ from Ref.~\onlinecite{Oguri2018a}, we evaluate the above expression and find that the sign of the low-temperature  thermopower is determined by the factor $s(B) = \sum_{\sigma}\sin(2\delta_{\sigma})\chi_{\sigma\sigma}(T=0)/\sum_{\sigma}\sin^2(\delta_{\sigma})$, where $\delta_{\sigma}=\pi n_{0\sigma}(0)$ is the spin $\sigma=\uparrow,\downarrow$
conduction electron phase shift, and $\chi_{\sigma\sigma}(T)$ is a static susceptibility defined by $\chi_{\sigma\sigma}(T)=\int_{0}^{1/T}d\tau \langle n_{0\sigma}(\tau)n_{0\sigma}(0)\rangle$. The latter quantities can be evaluated essentially exactly within the NRG for arbitrary $B$, and thereby allow $B_0$ to be extracted via $s(B=B_{0})=0$. A comparison between the NRG and the Fermi-liquid results for $B_{0}$ vs ${\rm v}_{g}$
is shown in Fig.~\ref{fig:fig2} for the case of $U/\Gamma=8$. The excellent agreement provides a check on the validity of the NRG results for $S$. 

\begin{figure}[t]
\centering 
\includegraphics[width=0.95\columnwidth]{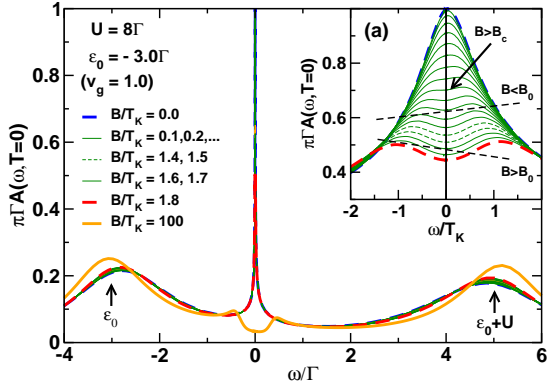}
\caption 
{Spectral function $\pi\Gamma A(\omega,T=0)$ vs $\omega/\Gamma$ for increasing $B/T_{\rm K}$ in the Kondo regime.
 Inset (a): Evolution of the low-energy Kondo resonance with increasing $B$, showing
  that it splits at a field $B_{c}\approx 0.75T_{\rm K}$ below the field $B_{0}\approx 1.45T_{\rm K}$ at which the slope of  $\pi\Gamma A(\omega,T=0)$ at the Fermi level [and hence $S(T\to 0)$] changes sign. 
  Dashed lines indicate the slope of $\pi\Gamma A(\omega,T=0)$ at the Fermi level for $B<B_{0}$ and $B>B_0$. 
  All parameters as in Fig.~\ref{fig:fig1}. 
}
\label{fig:fig3}
\end{figure}
Experiments on quantum dots often measure the field $B_c\approx 0.75 T_{\rm K}$ at which the Kondo resonance splits in a magnetic field [local minimum in $A(\omega,T=0)$ with positive
second derivative at $\omega=0$]. In contrast, $B_0$ represents the field at which $S(T\to 0)$ changes sign [resulting from a change in sign of the slope of $A(\omega,T=0)$ at $\omega=0$; see Eq.~(\ref{eq:sommerfeld})]. How do $B_c$ and $B_0$ compare ?
We find that $B_0$, while being of order $T_{\rm K}$, is generally larger than the field $B_c\approx 0.75 T_{\rm K}$ at which the Kondo resonance splits in a magnetic field [see Fig.~\ref{fig:fig2}(a)].
To illustrate this, we show in Fig.~\ref{fig:fig3} the $T=0$ NRG spectral function for increasing $B$ values \cite{Bulla1998, Hofstetter2000,Peters2006,Weichselbaum2007}.
The detailed behavior of the low-energy Kondo resonance in a magnetic field in Fig.~\ref{fig:fig3}(a) shows that the  value of
$B_{0}$ exceeds $B_{c}\approx 0.75 T_{\rm K}$ \footnote{This value, $B_c/T_{\rm K}\approx 0.75$, found in recent works \cite{Oguri2018a,Oguri2018b,Filippone2018}, agrees with Ref.~\onlinecite{Costi2000}, which found $B_c/T_{\rm K}^{\rm HWHM}\approx 0.50$ where $T_{\rm K}^{\rm HWHM}$ is the HWHM of the $T=0$ Kondo resonance given by 
$T_{\rm K}^{\rm HWHM}/T_{\rm K}\approx 1.5$, resulting in $B_c/T_{\rm K}\equiv (B_c/T_{\rm K}^{\rm HWHM})(T_{\rm K}^{\rm HWHM}/T_{\rm K})=0.5\times 1.5=0.75$.
}. For $U/\Gamma\gg 1$, we see  from Fig.~\ref{fig:fig2}(a) that $B_{0}$ can be up to twice as large as
$B_{c}$, whereas for $U/\Gamma=3.2$, relevant to the experiment of Ref.~\onlinecite{Svilans2018}, $B_0$ can be up to
$30\%-40\%$ larger than $B_{c}$. 
\begin{figure}[t]
\centering 
  \includegraphics[width=0.95\columnwidth]{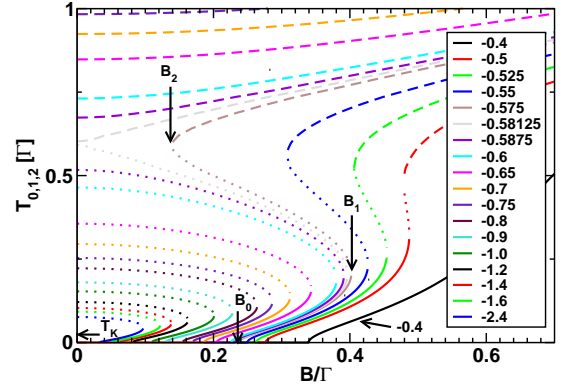}
  \caption{$T_{i=0,1,2}(B)$ vs $B/\Gamma$ for $U/\Gamma=5$ and different $\varepsilon_0/\Gamma$, 
    listed in the legend, and ranging from $\varepsilon_0/\Gamma=-2.4$ (${\rm v}_g=0.1$), in the Kondo regime, to $\varepsilon_0/\Gamma=-0.4$ (${\rm v}_g=2.1$), in the mixed valence regime. 
    Solid lines: $T_{0}(B)$; dotted lines: $T_1(B)$; dashed lines: $T_2(B)$. Values of $B_0,B_1$ and $B_2$ are
    indicated for $\varepsilon_0/\Gamma =-0.575$ in the ``weak Kondo regime''. Horizontal arrow: midvalley $T_{\rm K}$.
 }
\label{fig:fig4}
\end{figure}

Finally, since the essential signatures of the Kondo effect in the thermopower are the sign changes at the temperatures $T_0(B), T_1(B)$ and $T_2(B)$, we present  in Fig.~\ref{fig:fig4}, for $U/\Gamma=5$, their detailed evolution with magnetic field and local level position (i.e., gate voltage) in  the Kondo regime, and, show how this evolution is modified upon entry to the mixed valence regime at $\varepsilon_0/\Gamma \approx -0.5 $. As outlined in the context of  Fig~\ref{fig:fig1}(a), deep in the Kondo regime, $T_0(B)$ [$T_1(B)$] increase (decrease) with increasing $B$ up until $B=B_1$ when $T_0(B)$ and $T_1(B)$ merge, while for $B>B_1$ only the sign change at $T_2(B)$ remains.
Outside the Kondo regime, 
$\varepsilon_0/\Gamma \gtrsim -0.5 $, the temperatures $T_0(B),T_1(B)$ and $T_2(B)$ coalesce into a single temperature $T_0(B)$. For a narrow range of level positions,
$-0.6 \lesssim \varepsilon_0/\Gamma \lesssim -0.5$ ($1.9 \lesssim {\rm v}_g\lesssim 2.1$), between the Kondo and mixed valence regimes, which we term the ``weak Kondo regime,'' a more complex evolution of $T_1(B)$ and $T_2(B)$ with $B$ is observed: $T_1(B)$ and $T_2(B)$ bifurcate from a common (gate-voltage dependent) value at $B=B_{2}$ with $0\leq B_2 \leq B_1$. The coalescence of  $T_0(B),T_1(B)$ and $T_2(B)$ to the single temperature $T_0(B)$, then occurs upon entry to the mixed valence regime $\varepsilon_0/\Gamma \approx -0.5$ when $B_{2}=B_1$. 

{\em Comparison with experiment.}\,As detailed elsewhere \cite{Costi2019b}, results such as those in Fig.~\ref{fig:fig4} allow the full gate-voltage dependence of the thermoelectric response of Kondo-correlated {\rm InAS} quantum dots at different magnetic fields \cite{Svilans2018} to be explained.
Here, we consider a simpler quantity, the slope of the linear response thermocurrent $I_{\rm th}/\Delta T \equiv G(T)S(T)$ at ${\rm v}_g=0$, i.e., $\sigma(T)=d [G(T)S(T)]/d{\rm v}_{g}|_{{\rm v}_g=0}$, where $G(T)$ is the electrical conductance \footnote{See Refs.~\onlinecite{Prete2019,Costi2019b} for the relation between the thermocurrent and thermopower.}, and compare this with corresponding results from Ref.~\onlinecite{Svilans2018}. Evidently, since $G(T)$ is always positive and symmetric in ${\rm v}_g$, $\sigma(T)$ exhibits the same Kondo-induced sign changes at  $T_0,T_1$ and $T_2$ as $S(T)$ at small finite ${\rm v}_g$ and thereby it suffices to determine the Kondo-induced signatures in the thermopower.

We focus on device QD1a of Ref.~\onlinecite{Svilans2018}, which has $U=3.5\, {\rm meV}$, and $\Gamma=1.1\, {\rm meV}$ ($U/\Gamma=3.2$), resulting in a midvalley 
$T_{\rm K}^{\rm exp}\equiv T_{\rm K1}\approx 1.0\, K$.
From the value of $\Gamma$, and the measured $g$ factor $g\approx 9$ for {\rm InAs} quantum dots \cite{Kretinin2011,Svilans2018}, we carry out calculations for $\sigma(T)$ vs $T/\Gamma$
at the experimental field values $B=0.0\,T ,0.5\,T, 1.0\,T$ and $2.0\,T$.

\begin{figure}[t]
\centering 
  \includegraphics[width=0.95\columnwidth]{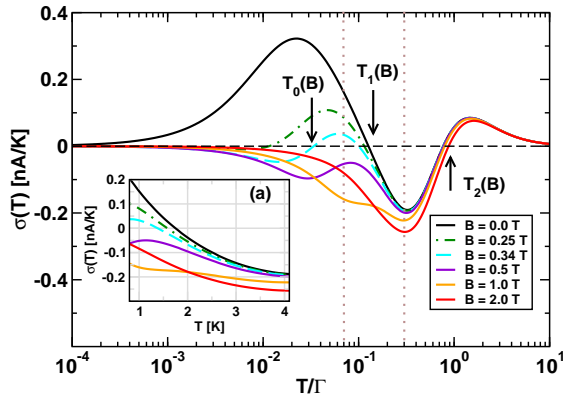}
  \caption{
Main panel: $\sigma(T)=d [G(T)S(T)]/d{\rm v}_{g}|_{{\rm v}_g=0}$ vs $T/\Gamma$ for the 
    four field values $B= 0\, T, 0.5\, T, 1.0\, T$ and $2.0\, T$ of the 
    experiment \cite{Svilans2018} and two additional field values $B = 0.25\, T$ and $0.34\, 
    T$ between $0 T$ and $0.5 T$. The vertical dotted lines denote the 
    lowest and highest temperatures of the experiment. Inset (a):  same as main panel, with temperature 
    in units of $K$ in the range ($1-4\, K$) of the experiment and 
    using a linear scale. For $B = 0\, T <B_{0}({\rm v}_g=0)\approx 0.23\, 
    T$ sign changes at $T_1(B)$ and $T_2(B)$ are seen, for $B=0.5\, T, 
    1.0\, T$ and $2.0\, T$ we have that $B > B_1({\rm v}_g=0)\approx 0.4\, 
    T$ and only the sign change at $T_2(B)\approx 10\, K$ is observable. 
  }
\label{fig:fig5}
\end{figure}

Figure \ref{fig:fig5} shows $\sigma(T)$ at the four experimental field values and at two additional ones at
$B=0.25\,T$ and $0.34\, T$ (to be discussed below) and over the whole temperature range. Figure \ref{fig:fig5}(a) restricts to the measured temperature window $1-4\,K$ and can be directly compared with
Fig.~4(f) of Ref.~\onlinecite{Svilans2018}. The possible sign changes in $\sigma(T)$, like in $G(T)S(T)$, can be
understood depending on whether  (i) $B<B_0$, (ii) $B_0< B < B_1$, or (iii)  $B>B_1$. The sign change
at $T_2(B)\approx 10\,K$ lies outside the measurement window, and will not be considered further \cite{Costi2019b}.
Starting with $B=0\, T\lesssim B_0({\rm v}_g=0)\approx 0.23\,T$ [Fig.~\ref{fig:fig2}(a)], we observe the expected sign change in $\sigma(T)$ upon increasing $T$ through $T_1(0)$, as seen also in experiment. For the cases $B = 1.0\, T$ and $2.0\ T$, we have $B>B_1({\rm v}_g=0)\approx 0.4\,T$, and in accordance with theory, no sign change is observed and none is found in experiment (within the measurement window). For the case $B=0.5\, T > B_1({\rm v}_g=0)\approx 0.4\,T$ [Fig.~\ref{fig:fig2}(b) and Ref.~\onlinecite{Costi2019b}], a qualitatively similar behavior of $\sigma(T)$ vs $T$ is observed in both theory and experiment (linear at higher $T$ with a leveling off at the lowest $T$), but in contrast to theory,
which does not predict a sign change of $\sigma(T)$ at this field value, the measurement finds a sign change, 
interpreted as $T_1(B)\approx 2.5\,K > T_1(0)\approx 1.8\,K$. The latter increase of $T_1(B)$ with $B$, however, is  inconsistent with theory, which predicts the opposite trend [see Fig.~\ref{fig:fig1}(b) and Fig.~\ref{fig:fig4}], and, inconsistent with measurements on all other devices in Ref.~\onlinecite{Svilans2018}, which agree with our predicted trend whenever a sign change at $T_1(B)$ is present.
To summarize, our comparison with experiment shows that the Kondo-induced sign change at $T_0(B)$ in $\sigma(T)$ has not been measured (in contrast to the one at $T_1$ for $B=0$). In order to observe the sign change at $T_0(B)$ for device QD1a, one needs to use smaller fields in the range $B_0\approx 0.23\,T \leq B \leq B_1\approx 0.4\,T$. Two fields, $B=0.25\,T$ and $0.34\,T$, satisfying this condition, and exhibiting the predicted sign change at
$T_0(B)\approx 0.16\,K$ and $0.45\,K$, respectively,  are shown in Fig.~\ref{fig:fig5}.

{\em Conclusions.}\,In summary, we investigated the magnetic field dependence of the thermopower of a Kondo-correlated quantum dot by applying the NRG technique  to the Anderson impurity model in a magnetic field. In the Kondo regime, we found, in addition to the known sign changes of $S(T)$ at $T_1$ and $T_2$, an  
additional sign change at a temperature $T_0(B)<T_{1}(B)<T_2(B)$ for magnetic fields $B$ exceeding a gate-voltage dependent value $B_0$. The  field $B_0$, of order $T_{K}$ in the Kondo regime, is comparable, but quantitatively different, to the field $B_c$ for the splitting of the Kondo resonance. Our results are of relevance in the light of recent advances in characterizing the thermoelectric properties of nanodevices \cite{Cui2017,Josefsson2018,Svilans2018,Dutta2019,Prete2019}. They explain, for example, the essential observations (Fig.~\ref{fig:fig5} and Ref.~\onlinecite{Costi2019b}) in the field dependence of the thermoelectric response of Kondo-correlated quantum dots \cite{Svilans2018}, and could serve as a guideline for interpreting future experiments on field-dependent thermoelectric transport through such systems.  
\begin{acknowledgments}
 We acknowledge discussions with L. Merker and supercomputer support by the John von Neumann institute for Computing (J\"ulich).
\end{acknowledgments}
\bibliography{mtep} 
\end{document}